\documentclass[8pt,twocolumn,letterpaper]{article}

\usepackage{cvpr}
\usepackage{times}
\usepackage{epsfig}
\usepackage{graphicx}
\usepackage{amsmath}
\usepackage{amssymb}
\usepackage{mathtools}
\usepackage{multirow} 
\usepackage{algorithm}
\usepackage{algpseudocode}
\usepackage{booktabs}

\usepackage{setspace}
\usepackage[breaklinks=true,bookmarks=false]{hyperref}

\cvprfinalcopy 

\def\cvprPaperID{****} 

\begin{document}

\title{Deviation Based Pooling Strategies For Full Reference Image Quality Assessment}

\author{Hossein Ziaei Nafchi $\qquad \qquad$ Hossein Ziaei Nafchi\\
\'Ecole de Technologie Sup\'erieure\\
Montreal, QC, Canada H3C 1K3\\
{\tt\small hossein.zi@synchromedia.ca}
}

\title{Deviation Based Pooling Strategies For Full Reference Image Quality Assessment}

\author{Hossein Ziaei Nafchi, Rachid Hedjam, Atena Shahkolaei and Mohamed Cheriet\\
\'Ecole de Technologie Sup\'erieure\\
\textit{ \normalsize ~\{hossein.zi@synchromedia.ca, ~rachid.hedjam@mcgill.ca, ~a.shahkolaei@yahoo.com, ~mohamed.cheriet@etsmtl.ca\}}}

\maketitle

\begin{abstract}
The state-of-the-art pooling strategies for perceptual image quality assessment (IQA) are based on the mean and the weighted mean. They are robust pooling strategies which usually provide a moderate to high performance for different IQAs. Recently, standard deviation (SD) pooling was also proposed. Although, this deviation pooling provides a very high performance for a few IQAs, its performance is lower than mean poolings for many other IQAs. In this paper, we propose to use the mean absolute deviation (MAD) and show that it is a more robust and accurate pooling strategy for a wider range of IQAs. In fact, MAD pooling has the advantages of both mean pooling and SD pooling. The joint computation and use of the MAD and SD pooling strategies is also considered in this paper. Experimental results provide useful information on the choice of the proper deviation pooling strategy for different IQA models.

\end{abstract}


%


\maketitle

\section{Introduction}
\label{sec:intro}


Automatic image quality assessment (IQA) plays a significant
role in many image processing applications. IQA is
commonly used for monitoring, benchmarking, image restoration and parameter optimization \cite{SSIM, datasets2012}. Full reference IQAs, which fall within the scope of this paper, evaluate the perceptual quality of a distorted image with respect to its reference image. IQAs mimic the average quality predictions of human observers. This is a non-trivial task because images may suffer from various types
and degrees of distortions. 

 Among IQAs, the mean squared error (MSE) is widely
used because of its simplicity. However, in many situations, it
does not correlate with human perception of image fidelity
and quality \cite{spm2009}. A number of popular and/or high performance IQAs are SSIM \cite{SSIM}, MS-SSIM \cite{MSSSIM}, VIF \cite{VIF}, VSNR \cite{VSNR}, MAD \cite{MAD}, FSIM \cite{FSIM}, GS \cite{GS}, GMSD \cite{GMSD}, VSI \cite{VSI}, and others \cite{Metrics2011}.  

Usually, IQAs measure local similarity and produce a similarity
score by a pooling strategy. This local quality
measurement can be performed on the image, different representations of the original image, or their combination. For example, SSIM and MSSSIM use statistics of smoothed source and distorted images. FSIM \cite{FSIM} uses a phase-derived map and another gradient-based map for quality assessment. GS \cite{GS} is a contrast and structure variant metric that utilizes specialized gradient magnitude and image contrast
of the image. GMSD \cite{GMSD} also utilizes gradient magnitude. Many of the available full reference IQAs follow this top-down architecture \cite{Metrics2011}. While average pooling \cite{SSIM, MSSSIM, FSIM, GS} and average weighted pooling \cite{pooling2009, IWSSIM, VSI} are widely used in the literature, GMSD uses standard deviation pooling \cite{GMSD}. For the purpose of this paper, we will take an overview of the FSIM \cite{FSIM} and GMSD \cite{GMSD} indices.

The feature similarity (FSIM) index \cite{FSIM} uses phase congruency ($PC$) \cite{kovesipc} as its main feature, and an image gradient magnitude $G$ as its secondary feature. Phase congruency similarity $S_{PC}$ and gradient magnitude similarity $S_G$ are calculated and then combined as $S_L(\textbf{x})=[S_{PC}(\textbf{x})]^\alpha[S_G(\textbf{x})]^\beta$, where $\textbf{x}$ is an image pixel. For the purpose of pooling, first, the maximum $PC$ of reference and distorted images $PC_m$ is computed. FSIM is then computed by the following mean pooling: 
\begin{equation}
  \ \text{FSIM}=\frac{\sum_{\textbf{x}} \big( S_L(\textbf{x})PC_m(\textbf{x}))}{\sum_{\textbf{x}}PC_m(\textbf{x} \big) }.
  \label{equ:1}
\end{equation}                    

FSIM is among the leading indices in the literature; however, the high computation cost of the phase congruency makes it an inefficient index.

GMSD uses the Prewitt operator to calculate gradient magnitudes of reference and distorted images, $G_\mathcal{R}$ and $G_\mathcal{D}$. From these, a gradient magnitude similarity (GMS) is calculated by:
\begin{equation}
  \ \text{GMS(\textbf{x})}=\frac{2G_\mathcal{R}(\textbf{x})G_\mathcal{D}(\textbf{x})+c}{G^2_\mathcal{R}(\textbf{x})+G^2_\mathcal{D}(\textbf{x})+c}.
  \label{equ:GMS}
\end{equation}   
where $c$ is a positive constant that supplies numerical stability. The GMSD is then calculated by a deviation pooling strategy, which is the standard deviation of GMS values. GMSD provides high performance for different datasets and is very efficient. 

The image gradients are sensitive to image distortions; different local structures in a distorted image suffer from different degrees of degradations \cite{GMSD}. This is the motivation that authors in \cite{GMSD} used to explore the standard variation of the gradient-based local similarity map for overall image quality prediction. The standard deviation is the square root of the mean of the squares of the individual deviations. A problem with standard deviation is that the larger deviations are overemphasized in the process of squaring the deviations, since taking the square root is not a complete reversal.    

In this paper, the underestimated mean absolute deviation pooling is proposed for IQAs as it is more tolerant of the large deviations. We show that the mean absolute deviation is a faster and more reliable pooling than the standard deviation for different IQAs. Also for some IQAs, a combination of these two pooling strategies results in indices that perform better. At the same time, the joint calculation of the standard deviation and the mean absolute deviation is still efficient.

\section{Deviation pooling strategies}
\label{pooling}

Deviation pooling for IQAs is rarely used in the literature, except the standard deviation used in GMSD \cite{GMSD}. Deviation is the variation of data values compared to a measure of central tendency (MCT) such as the mean, median, or mode. A deviation can be seen as the Minkowski distance of order $\rho$ between vector \textbf{x} and a MCT:
\begin{equation}
  \ \text{D}(\textbf{x},\text{MCT})^\rho = \Big(\sum_{i=1}^{N}\big| \textbf{x}_i - \text{MCT} \big|^\rho \Big)^{1/\rho}.
  \label{equ:minkowski}
\end{equation}                    
where, $\rho \geq 1$ indicates the type of deviation. For the purpose of this paper, it can be seen as:
\begin{equation}
  \ \widehat{\text{D}}(\textbf{x},\text{MCT})^\rho = \Big( \frac{1}{N} \sum_{i=1}^{N}\big| \textbf{x}_i - \text{MCT} \big|^\rho \Big)^{1/\rho}.
  \label{equ:minkowski1}
\end{equation}                    

Since above equation includes a MCT, it is different than the Minkowski pooling \cite{Mpooling2006}, and the Minkowski metric \cite{handbook2000}. In the following, the standard deviation pooling strategy introduced in GMSD is revisited. We then suggest using the mean absolute deviation pooling and show that these two pooling strategies can be jointly calculated.

\subsection{Standard deviation ($\mathcal{SD}$) pooling}
\label{sdpooling}

Standard deviation is a simple and very common statistical measure of the spread of scores within a set of data. Let LS$^\mathcal{M}$ denote the mean of an arbitrary local similarity (LS) map computed by an IQA model. Its standard deviation can be computed by:
\begin{equation}
  \ \text{LS}^ \mathcal{SD} = \sqrt{\frac{1}{N}\sum_{i=1}^N \big(LS(i)-LS^\mathcal{M}\big)^2}.
  \label{equ:std1}
\end{equation}                    
which is equal to equation (\ref{equ:minkowski1}) when $\rho=2$.

\subsection{Mean absolute deviation ($\mathcal{MAD}$) pooling}
\label{madpooling}

Given the LS$^\mathcal{M}$, the mean absolute deviation LS$^\mathcal{MAD}$ of an LS map is calculated by:
\begin{equation}
  \ \text{LS}^\mathcal{MAD}=\frac{1}{N}\sum_{i=1}^N \big| LS(i)-LS^\mathcal{M} \big|.
  \label{equ:mad1}
\end{equation}                    
which is also equal to equation (\ref{equ:minkowski1}) when $\rho=1$. In the experimental results section, the performance of the $\mathcal{MAD}$ pooling based indices is evaluated. 

\subsection{Double deviation ($\mathcal{DD}$) pooling}
\label{madpooling}

We also found that a combination of the $\mathcal{SD}$ and $\mathcal{MAD}$ pooling strategies provides higher performance than either LS$^\mathcal{SD}$ or LS$^\mathcal{MAD}$ for some IQAs. Furthermore, these two poolings can be computed at the same time using the following formulas: 
\begin{equation}
  \ {D_i}=\big| LS(i)-LS^ \mathcal{M} \big| ,
  \label{equ:di}
\end{equation}                    
\begin{equation}
  \ \text{LS}^\mathcal{MAD}=\frac{1}{N} \sum_{i=1}^N D_i.
  \label{equ:mad2}
\end{equation}                    
\begin{equation}
  \ \text{LS}^\mathcal{SD}=\sqrt{ \frac{1}{N} \sum_{i=1}^N D^2_i }.
  \label{equ:sd2}
\end{equation}                    

The joint computation of the LS$^\mathcal{SD}$ and LS$^\mathcal{MAD}$ is likewise computationally efficient. We call this combination the double deviation $\mathcal{DD}$ pooling strategy and compute it by:
\begin{equation}
  \ \text{LS}^\mathcal{DD} = \alpha ~ \text{LS}^\mathcal{SD} + (1-\alpha) ~ \text{LS}^\mathcal{MAD} ~.
  \label{equ:fd}
\end{equation}                    
where $0 \le \alpha \le 1$ adjusts the relative importance of the
LS$^\mathcal{SD}$ and LS$^\mathcal{MAD}$ indices. We are able to combine these two pooling strategies because they share almost the same statistical characteristics and values. The usefulness of this combination is verified in the experiments. It should be noted that the median absolute deviation does not provide satisfactory performance, hence it is not evaluated in this paper.

\section{Experimental results}
\label{experiments}

In the experiments, four standard datasets are used. The LIVE dataset \cite{LIVEweb} contains 29 reference images and 779 distorted images of five categories. The TID 2008 \cite{TID2008} dataset contains 25 reference images and 1700 distorted images. For each reference image, 17 types of distortions of 4 degrees are available. CSIQ \cite{MAD} is another dataset that consists of 30 reference images; each is distorted using six different types of distortions at four to five levels of distortion. We also used the TID 2013 \cite{TID2013} dataset, which contains 25 reference images and 3000 distorted images. For each reference image, 24 types of distortions of 5 degrees are available. Also, three popular evaluation metrics were used in the experiments: the Spearman Rank order Correlation coefficient (SRC), the Pearson linear Correlation Coefficient (PCC), and the Root Mean Square Error (RMSE).

\begin{figure}[htb]
\small
\begin{minipage}[b]{0.99\linewidth}
  \centering
  \centerline{\includegraphics[height=6cm]{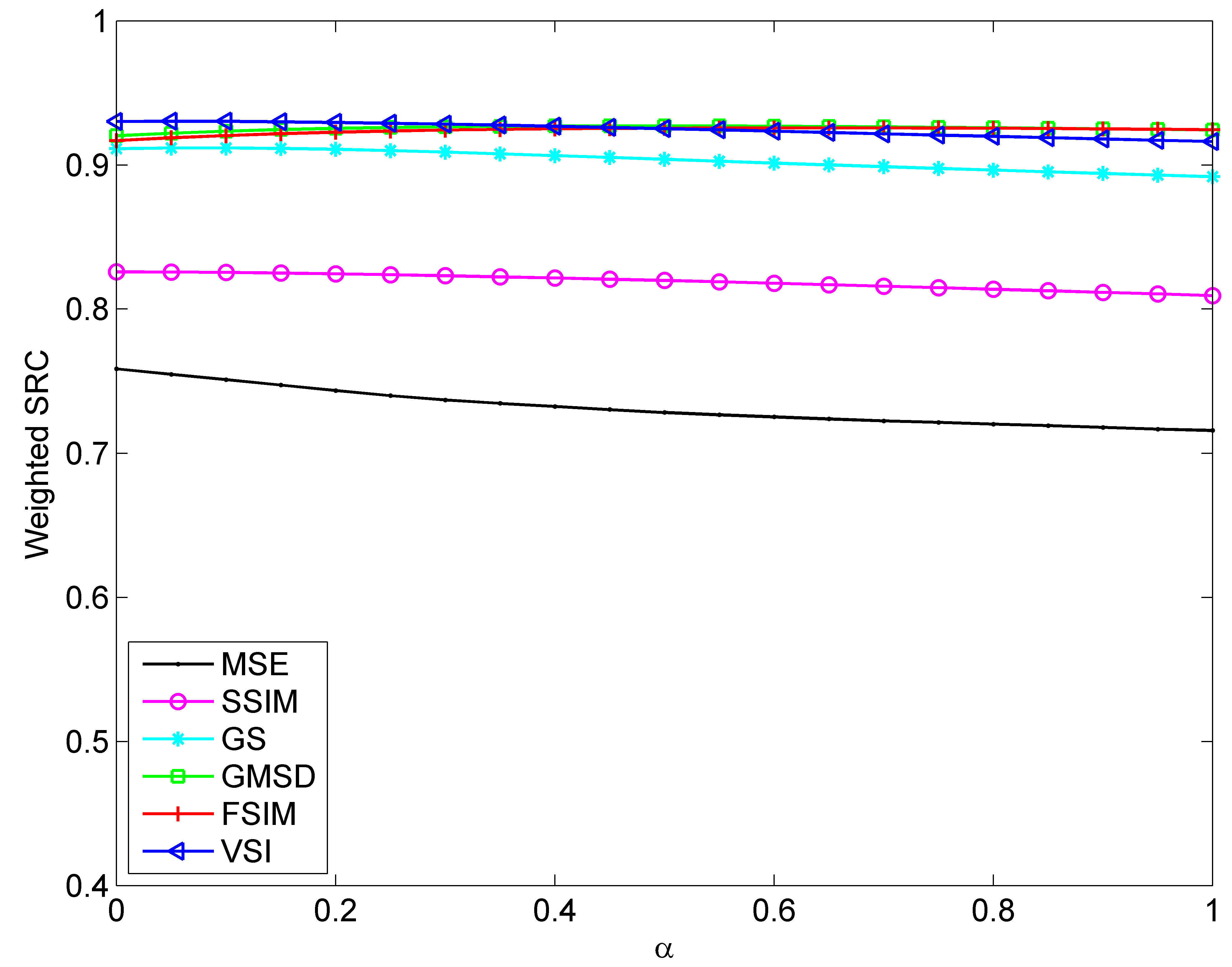}}
\end{minipage}
\caption{The impact of the parameter $\alpha$ on the weighted average SRC (Table \ref{results1}) performance of the double deviation $\mathcal{DD}$ pooling strategy.}
\label{fig:chart1}
\end{figure}

\begin{figure}[htb]
\small
\begin{minipage}[b]{0.99\linewidth}
  \centering
  \centerline{\includegraphics[height=6cm]{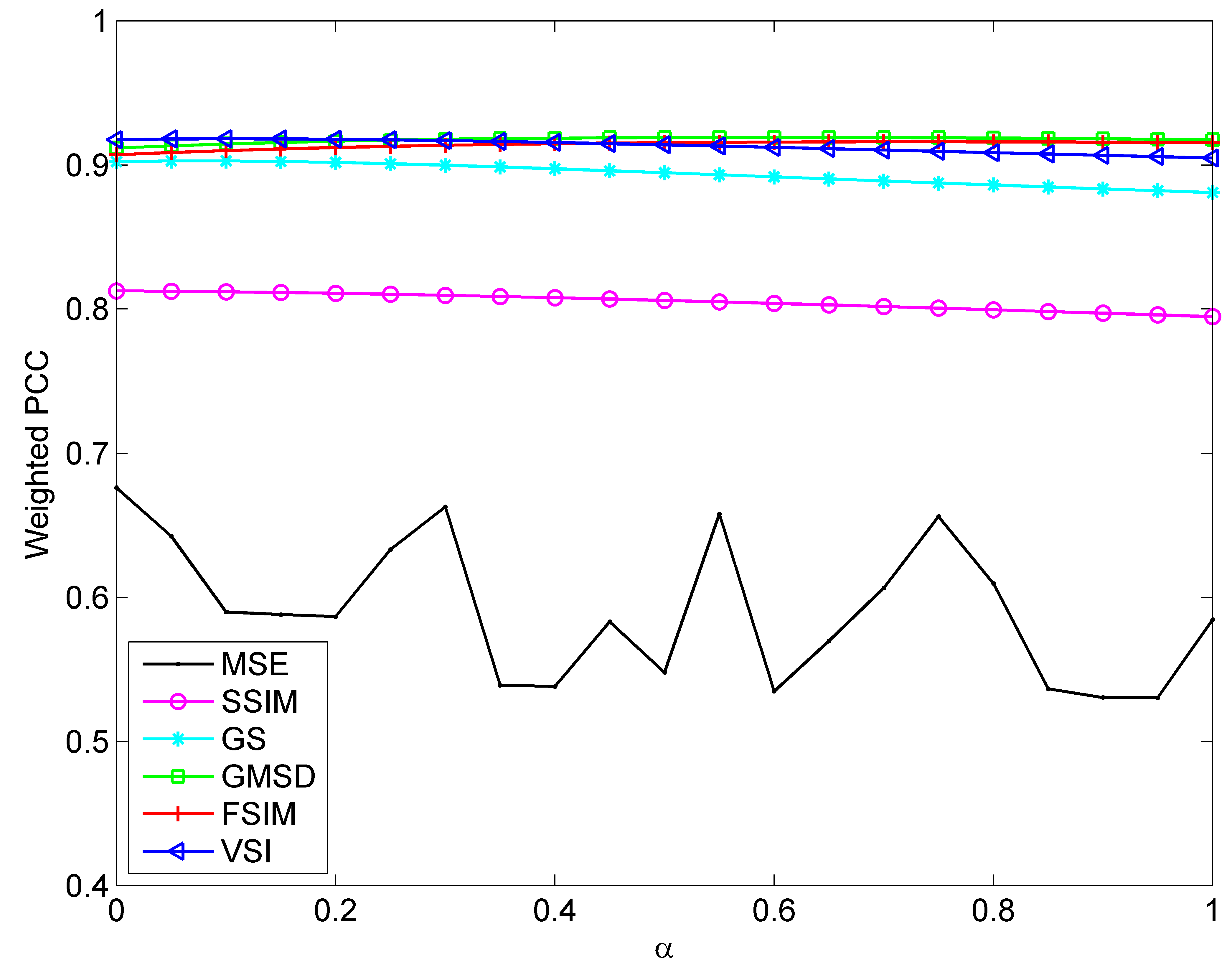}}
\end{minipage}
\caption{The impact of the parameter $\alpha$ on the weighted average PCC (Table \ref{results1}) performance of the double deviation $\mathcal{DD}$ pooling strategy. Note the MSE curve which is nonlinear.}
\label{fig:chart11}
\end{figure}

MSE, SSIM \cite{SSIM}, GS \cite{GS, GSweb}, FSIM \cite{FSIM, FSIMweb}, GMSD \cite{GMSD, GMSDweb}, and VSI \cite{VSI} were used in the experiments using two to three deviation pooling strategies\footnote{Code: https://dl.dropboxusercontent.com/u/74505502/DeviationPoolings.m}. Tables \ref{results1} and \ref{results2} provide a performance comparison between different indices using different pooling strategies. In reference to Fig. \ref{fig:chart1}, if $\mathcal{DD}$ pooling provides higher performance than the others, its performance is added to Tables \ref{results1} and \ref{results2}. Also, some of the state-of-the-art indices are added to the end of Tables \ref{results1} and \ref{results2}. In Figs. \ref{fig:chart1} and \ref{fig:chart11}, $\alpha=0$ postulates to the $\mathcal{MAD}$ pooling, and $\alpha=1$ postulates to the $\mathcal{SD}$ pooling. In fact, Figs. \ref{fig:chart1} and \ref{fig:chart11} show SRC and PCC performance variations of different indices utilizing the $\mathcal{DD}$ pooling.

\begin{figure*}[htb]
\scriptsize
\begin{minipage}[b]{0.33\linewidth}
  \centering
  \centerline{\includegraphics[height=3.9cm]{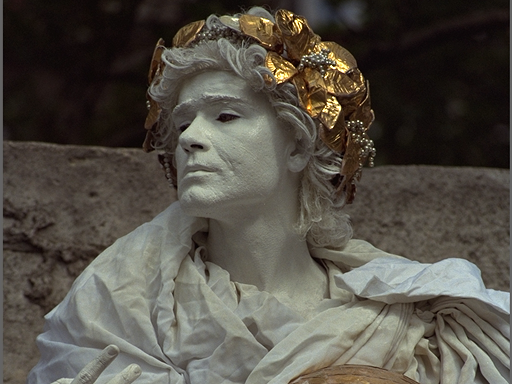}}
\centerline{(a) original image}
\centerline{taken from TID 2008 \cite{TID2008}}
\centerline{}
\end{minipage}
\begin{minipage}[b]{.33\linewidth}
  \centering
  \centerline{\includegraphics[height=3.9cm]{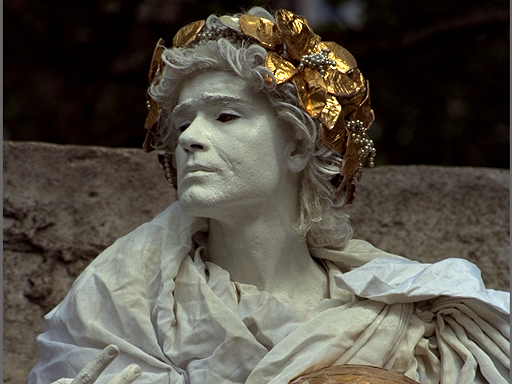}}
\centerline{(b) subjective score = 7.1290}
\centerline{$\mathcal{SD}$: GS = 0.0030, GMS = 0.0225}
\centerline{$\mathcal{MAD}$: GS = 0.0013, GMS  = 0.0075}
\end{minipage}
\begin{minipage}[b]{0.33\linewidth}
  \centering
  \centerline{\includegraphics[height=3.9cm]{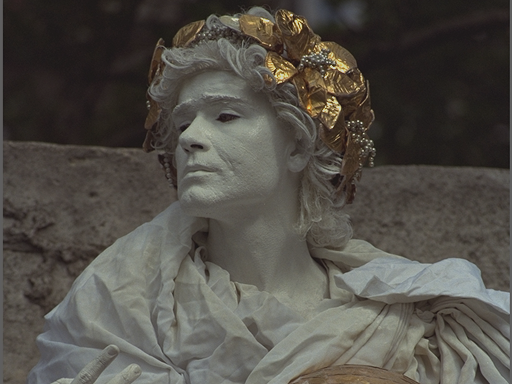}}
\centerline{(c) subjective score = 5.0645}
\centerline{$\mathcal{SD}$: GS = 0.0027, GMS = 0.0127}
\centerline{$\mathcal{MAD}$: GS = 0.0020, GMS  = 0.0111}
\end{minipage}
\caption{A comparison of $\mathcal{SD}$ and $\mathcal{MAD}$ for contrast distortion type. (a) original image, (b)-(c) contrast distortion images of (a). Note that higher subjective scores and lower GS/GMS scores indicate higher quality. Clearly, GS/GMS $\mathcal{MAD}$ provides better judgment than GS/GMS $\mathcal{SD}$.}
\label{fig:out1}
\end{figure*}

\begin{table*}[!t]
\centering
\caption{Performance comparison of the different quality indices with and without deviation pooling strategies on LIVE \cite{LIVEweb}, CSIQ \cite{MAD} and TID 2008 \cite{TID2008} datasets.}
\scriptsize
\begin{tabular}{c|ccc|ccc|ccc|cc|ccl}
\hline
\multirow{2}{*}{IQA} & \multicolumn{3}{c|}{LIVE (779 images)}               & \multicolumn{3}{c|}{CSIQ (886 images)}              & \multicolumn{3}{c|}{TID2008 (1700 images)}          & \multicolumn{2}{c|}{Weighted avg} & \multicolumn{3}{c}{SRC (distortions)}                       \\ \cline{2-15} 
                     & SRC             & PCC             & RMSE             & SRC             & PCC             & RMSE            & SRC             & PCC             & RMSE            & SRC             & PCC             & avg             & min             & \multicolumn{1}{c}{std} \\ \hline\hline
MSE                  & 0.8756          & \textbf{0.8723} & \textbf{13.3597} & 0.8058          & \textbf{0.7512} & \textbf{0.1733} & 0.5531          & 0.5734          & 1.0994          & 0.6943          & \textbf{0.6894} & \textbf{0.8473} & \textbf{0.5815} & \textbf{0.1078}         \\
MSE $\mathcal{SD}$ \cite{GMSD}              & \textbf{0.8771} & 0.5707          & N/A       & \textbf{0.8344} & 0.6448          & 0.2007          & 0.5801          & 0.5593          & 1.1124          & 0.7158          & 0.5845          & 0.8290          & 0.2624          & 0.1658                  \\
MSE $\mathcal{MAD}$              & 0.8746          & 0.8716          & 13.3944          & 0.8239          & 0.5594          & 0.2176          & \textbf{0.6712} & \textbf{0.6473} & \textbf{1.0229} & \textbf{0.7585} & 0.6761          & 0.8422          & 0.5045          & 0.1300                  \\ \hline
SSIM \cite{SSIM}                & \textbf{0.9479} & \textbf{0.9449} & \textbf{8.9455}  & \textbf{0.8756} & \textbf{0.8613} & \textbf{0.1334} & 0.7749          & \textbf{0.7732} & \textbf{0.8511} & \textbf{0.8415} & \textbf{0.8361} & \textbf{0.8644} & \textbf{0.5246} & \textbf{0.1069}         \\
SSIM $\mathcal{SD}$ \cite{GMSD}             & 0.9174          & 0.9032          & 11.7261          & 0.8169          & 0.8094          & 0.1542          & 0.7560          & 0.7374          & 0.9063          & 0.8094          & 0.7948          & 0.8154          & 0.0083          & 0.2209                  \\
SSIM $\mathcal{MAD}$             & 0.9166          & 0.9017          & 11.8156          & 0.8388          & 0.8316          & 0.1458          & \textbf{0.7775} & 0.7619          & 0.8691          & 0.8258          & 0.8126          & 0.8255          & 0.2059          & 0.1836                  \\ \hline
GS \cite{GS}                  & \textbf{0.9561} & \textbf{0.9512} & \textbf{8.4327}  & 0.9108          & 0.8964          & 0.1164          & 0.8504          & 0.8422          & 0.7235          & 0.8908          & 0.8817          & \textbf{0.8915} & \textbf{0.6691} & \textbf{0.0923}         \\
GS $\mathcal{SD}$                & 0.9464          & 0.9409          & 9.2559           & \textbf{0.9308} & \textbf{0.9241} & \textbf{0.1003} & 0.8466          & 0.8307          & 0.7470          & 0.8919          & 0.8808          & 0.8872          & 0.6146          & 0.1108                  \\
GS $\mathcal{MAD}$               & 0.9538          & 0.9486          & 8.6494           & 0.9295          & 0.9198          & 0.1030          & \textbf{0.8823} & \textbf{0.8720} & \textbf{0.6568} & \textbf{0.9113} & \textbf{0.9023} & 0.8899          & 0.6523          & 0.1025                   \\ \hline
FSIM \cite{FSIM}               & \textbf{0.9634} & \textbf{0.9597} & \textbf{7.6780}  & 0.9242          & 0.9120          & 0.1077          & 0.8805          & 0.8738          & 0.6525          & 0.9112          & 0.9038          & 0.8881          & \textbf{0.6481} & \textbf{0.0934}         \\
FSIM $\mathcal{SD}$ \cite{GMSD}             & 0.9602          & 0.9579          & 7.8442           & \textbf{0.9566} & \textbf{0.9534} & \textbf{0.0792} & 0.8914 & 0.8762 & 0.6467 & 0.9245 & \textbf{0.9154} & 0.8843          & 0.4412          & 0.1268                  \\
FSIM $\mathcal{MAD}$             & 0.9609          & 0.9580          & 7.8370           & 0.9525          & 0.9460          & 0.0851          & 0.8783          & 0.8634          & 0.6770          & 0.9170          & 0.9071          & \textbf{0.8916} & 0.6195          & 0.1043 \\ 
FSIM $\mathcal{DD}$ ($\alpha=0.5$)            & 0.9611          & 0.9584          & 7.8003           & 0.9555          & 0.9507          & 0.0814          & \textbf{0.8935}          & \textbf{0.8775}          & \textbf{0.6436}          & \textbf{0.9255}          & \textbf{0.9155}          & 0.8882 & 0.5066          & 0.1164 \\ \hline
GMS $\mathcal{M}$                & 0.9595          & 0.9556          & 8.0489           & 0.9290          & 0.9127          & 0.1073          & 0.8477          & 0.8366          & 0.7351          & 0.8950          & 0.8842          & 0.8924          & \textbf{0.6301} & \textbf{0.0962}         \\
GMSD \cite{GMSD}              & 0.9603          & 0.9603          & 7.6214           & \textbf{0.9570} & \textbf{0.9541} & \textbf{0.0786} & 0.8907          & 0.8788          & 0.6404          & 0.9243          & 0.9175          & 0.8849          & 0.4659          & 0.1246                  \\
GMS $\mathcal{MAD}$              & \textbf{0.9627} & \textbf{0.9618} & \textbf{7.4802}  & 0.9532          & 0.9457          & 0.0853          & 0.8837          & 0.8711          & 0.6589          & 0.9203          & 0.9118          & \textbf{0.8935} & 0.6224          & 0.1034                  \\
GMS $\mathcal{DD}$ ($\alpha=0.5$)       & 0.9619          & 0.9614          & 7.5149           & 0.9559          & 0.9509          & 0.0813          & \textbf{0.8961} & \textbf{0.8830} & \textbf{0.6300} & \textbf{0.9271} & \textbf{0.9190} & 0.8889          & 0.5204          & 0.1160                  \\ \hline

VSI \cite{VSI}       & 0.9524          & 0.9482          & 8.6817           & 0.9423          & 0.9279          & 0.0979          &  0.8979 & 0.8762 & 0.6466 & 0.9222 & 0.9065 & 0.8987          & 0.6295          & 0.1036 
\\

VSI $\mathcal{SD}$ \cite{GMSD}              & 0.9546          & 0.9519          & 8.3757           & \textbf{0.9569} & \textbf{0.9532} & \textbf{0.0801} & 0.8775          & 0.8585          & 0.6881          & 0.9163          & 0.9048          & 0.8957          & 0.5393          & 0.1080                  \\

VSI $\mathcal{MAD}$              & \textbf{0.9577} & \textbf{0.9540} & \textbf{8.1952}  & 0.9546          & 0.9449          & 0.0859          & \textbf{0.9048}          & 0.8866         & 0.6207          & \textbf{0.9302}          & 0.9175          & \textbf{0.9027} & \textbf{0.6312}          & \textbf{0.0949} \\

VSI $\mathcal{DD}$ ($\alpha=0.1$)       & \textbf{0.9575}          & \textbf{0.9539}          & \textbf{8.1958}          & 0.9553          & 0.9462          & 0.0849          & \textbf{0.9048} & \textbf{0.8872} & \textbf{0.6193} & \textbf{0.9303} & \textbf{0.9182} & \textbf{0.9029} & 0.6275          & \textbf{0.0948}                  \\ \hline

MSSSIM \cite{MSSSIM}              & 0.9513 & 0.9489 & 8.6188  & 0.9133 & 0.8991 & 0.1149 & 0.8542 & 0.8451 & 0.7173 & 0.8922 & 0.8834 & 0.8796 & \textbf{0.6381} & \textbf{0.0993}         \\
VIF \cite{VIF}       & 0.9636          & 0.9604          & 7.6137           & 0.9195          & 0.9277          & 0.0980          &  0.7491 & 0.8084 & 0.7899 & 0.8436 & 0.8750 &  0.8949          & 0.5102          & \textbf{0.0987}                 \\
IWSSIM \cite{IWSSIM}       & 0.9567          & 0.9522          & 8.3472           & 0.9213          & 0.9144          & 0.1063          &  \textbf{0.8559} & \textbf{0.8579} & \textbf{0.6895} & \textbf{0.8965} & \textbf{0.8946} & 0.8708          & \textbf{0.6301}          & 0.1063                  \\ 
MAD \cite{MAD}       &  \textbf{0.9669}          & \textbf{0.9675}          & \textbf{6.9073}           &  \textbf{0.9466}          & \textbf{0.9500} & \textbf{0.0820}          &  0.8340 & 0.8290 & 0.7505 & 0.8944 & 0.8929 & 0.8387          & 0.0650          & 0.2152        \\ \hline
\end{tabular}
\label{results1}
\end{table*}

\begin{table}[!t]
\centering
\caption{Performance comparison of the different quality indices with and without deviation pooling strategies on TID 2013 dataset \cite{TID2013}.}
\scriptsize
\begin{tabular}{c|cc|ccc}
\hline
\multirow{2}{*}{IQA} & \multicolumn{2}{c|}{TID2013}      & \multicolumn{3}{c}{SRC (distortions)}                                   \\ \cline{2-6} 
                     & SRC             & PCC             & avg             & min             & std                                 \\ \hline \hline
MSE                  & 0.6394          & 0.4785          & \textbf{0.7951} & 0.0766          & \multicolumn{1}{l}{0.2355}          \\
MSE $\mathcal{SD}$               & 0.6164          & 0.5118          & 0.7899          & \textbf{0.0952} & \multicolumn{1}{l}{\textbf{0.2282}} \\
MSE $\mathcal{MAD}$              & \textbf{0.6891} & \textbf{0.7079} & 0.7897          & 0.0852          & \multicolumn{1}{l}{0.2413}          \\ \hline
SSIM  \cite{SSIM}               & 0.7417          & \textbf{0.7895} & \textbf{0.8075} & \textbf{0.3775} & \multicolumn{1}{l}{\textbf{0.1521}} \\
SSIM $\mathcal{SD}$              & 0.7292          & 0.7602          & 0.7993          & 0.0045          & \multicolumn{1}{l}{0.2257}          \\
SSIM $\mathcal{MAD}$             & \textbf{0.7463} & 0.7806          & 0.8041          & 0.1471          & \multicolumn{1}{l}{0.1989}          \\ \hline
GS \cite{GS}                  & 0.7946          & 0.8464          & \textbf{0.8351} & \textbf{0.3578} & \multicolumn{1}{l}{\textbf{0.1511}} \\
GS $\mathcal{SD}$                & 0.7801          & 0.8232          & 0.8284          & 0.3139          & \multicolumn{1}{l}{0.1731}          \\
GS $\mathcal{MAD}$               & \textbf{0.8081} & \textbf{0.8613} & 0.8332          & 0.3344          & \multicolumn{1}{l}{0.1584}          \\ \hline
FSIM \cite{FSIM}                & 0.8015          & 0.8589          & 0.8219          & \textbf{0.2748} & \textbf{0.1662}                     \\
FSIM $\mathcal{SD}$ \cite{GMSD}             & 0.8077          & 0.8602          & 0.8263          & 0.2126          & 0.1940                              \\
FSIM $\mathcal{MAD}$             & 0.8051          & 0.8547          & \textbf{0.8327} & 0.2691          & 0.1706                              \\
FSIM $\mathcal{DD}$ ($\alpha=0.5$)      & \textbf{0.8118} & \textbf{0.8614} & 0.8293          & 0.2304          & 0.1852                              \\ \hline
GMS $\mathcal{M}$                & 0.7884          & 0.8395          & \textbf{0.8370} & \textbf{0.3700} & \textbf{0.1500}                     \\
GMSD \cite{GMSD}              & 0.8044          & 0.8590          & 0.8300          & 0.2948          & 0.1815                              \\
GMS $\mathcal{MAD}$              & 0.8084          & 0.8608          & \textbf{0.8369} & 0.3450          & 0.1597                              \\
GMS $\mathcal{DD}$ ($\alpha=0.5$)       & \textbf{0.8111} & \textbf{0.8658} & 0.8332          & 0.3132          & 0.1729                              \\ \hline
VSI \cite{VSI}                 & \textbf{0.8965} & \textbf{0.9000} & 0.8514          & 0.1713          & 0.1787                              \\
VSI $\mathcal{SD}$               & 0.8556          & 0.8651          & 0.8633          & 0.3935          & 0.1274                              \\
VSI $\mathcal{MAD}$              & 0.8853          & \textbf{0.8965} & \textbf{0.8675} & \textbf{0.4564} & \textbf{0.1193}                     \\
VSI $\mathcal{DD}$ ($\alpha=0.1$)       & 0.8830          & 0.8951          & \textbf{0.8680} & \textbf{0.4542} & \textbf{0.1183}                     \\ \hline
MSSSIM  \cite{MSSSIM}             & \textbf{0.7859} & \textbf{0.8329} & 0.8109 & \textbf{0.4099} & \multicolumn{1}{l}{0.1506} \\
VIF \cite{VIF}                 & 0.6769          & 0.7720          & \textbf{0.8267}          & 0.3099          & \textbf{0.1464}                              \\
IWSSIM \cite{IWSSIM}               & 0.7779          & 0.8319          & 0.7978          & 0.3717          & 0.1601                              \\
MAD \cite{MAD}                 & 0.7807          & 0.8267          & 0.7556          & 0.0575          & 0.2644                              \\ \hline
\end{tabular}
\label{results2}
\end{table}

From Tables \ref{results1} and \ref{results2}, it can be seen that MSE $\mathcal{MAD}$ performs better than the MSE, while the MSE $\mathcal{SD}$ shows the lowest performance among them. The original SSIM outperforms its deviation-based versions. The $\mathcal{MAD}$ pooling has higher performance than the $\mathcal{SD}$ pooling for SSIM. For SSIM, the deviation poolings show very low performance for some of the distortions, while the $\mathcal{MAD}$ pooling is still more robust than the $\mathcal{SD}$ pooling. 

For GS, the $\mathcal{MAD}$ pooling outperforms the others. $\mathcal{SD}$ pooling for GS does not provide a high performance because GS uses image contrast, and we had already observed that $\mathcal{SD}$ pooling is not a good choice for MSE. While FSIM $\mathcal{SD}$ shows overall higher performance than FSIM $\mathcal{MAD}$, its performance for some distortion types is low. For FSIM, $\mathcal{DD}$ pooling with $\alpha=0.5$ provides higher overall performance than $\mathcal{SD}$ pooling. At the same time, $\mathcal{DD}$ pooling shows better quality prediction on distortion types than $\mathcal{SD}$ pooling. The overall performance of the GMS $\mathcal{MAD}$ and GMS $\mathcal{SD}$ indices is competitive; however, GMS $\mathcal{MAD}$ shows better quality prediction on the distortion types. The overall performance of GMS $\mathcal{DD}$ and its performance for distortion types are simultaneously higher than GMS $\mathcal{SD}$.

VSI is a high performance similarity index that was recently proposed in \cite{VSI}. Using $\mathcal{MAD}$ pooling, its performance improved considerably on the first three datasets for all of the measures used in this paper. For the TID 2013 dataset, however, its overall performance decreased by 1.2493\% for SRC and by 0.3889\% for PCC metrics. In turn, it has 1.8910\% better average prediction on distortion types. Also, the minimum quality prediction of VSI improved from 0.1713 to the 0.4564. These advantages show that $\mathcal{MAD}$ pooling is a good choice for VSI. $\mathcal{SD}$ pooling shows the lowest performance for VSI.

Overall, $\mathcal{MAD}$ pooling is more robust than $\mathcal{SD}$ pooling, especially for assessment of individual distortion types. The low min SRC values for $\mathcal{SD}$ pooling in Table \ref{results1} show its unreliability in comparison to $\mathcal{MAD}$ pooling. In general, higher orders of $\rho$ in equation (\ref{equ:minkowski1}) result in a worst and unstable assessment for distortion types. In other words, the std value in the last column of the Table \ref{results1} increases by increasing the $\rho$ value. It is worth noting that this fact may not always be true.

\begin{figure}[htb]
\small
\begin{minipage}[b]{0.99\linewidth}
  \centering
  \centerline{\includegraphics[height=6cm]{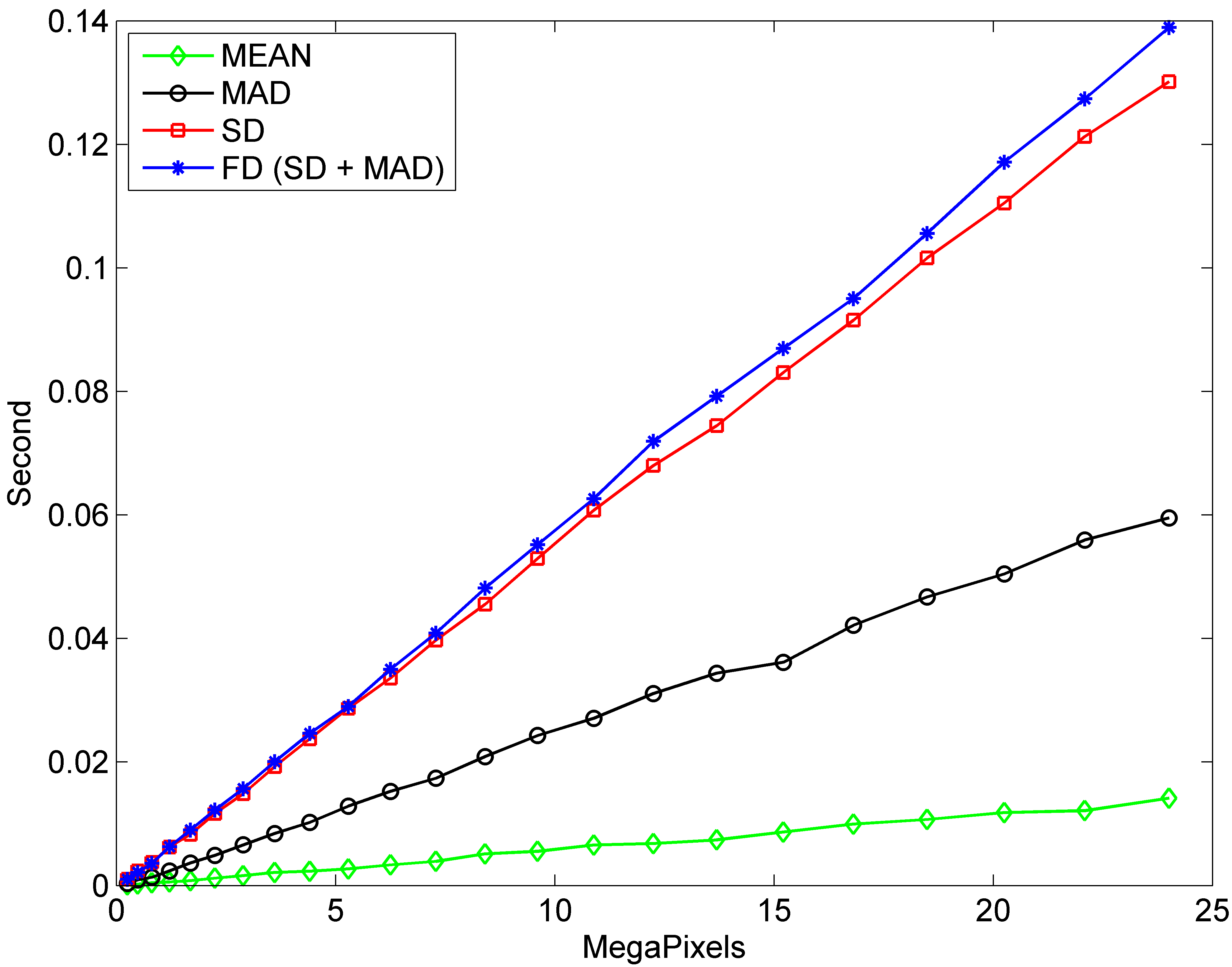}}
\end{minipage}
\caption{Run time versus the local similarity (LS) size of the mean pooling and three deviation pooling strategies used in this paper.}
\label{fig:chart2}
\end{figure}

Fig. \ref{fig:out1} shows an example in which $\mathcal{SD}$ pooling fails in assessment, while $\mathcal{MAD}$ pooling provides a true assessment for both the GS and GMS indices. Fig. \ref{fig:chart2} shows the run time of the three pooling strategies used in this paper. Our experiments were performed on a Core i7 3.4 GHz CPU with 16 GB of RAM running on MATLAB 2013b and Windows 7. $\mathcal{MAD}$ is the second fastest after the mean, while the joint calculation of $\mathcal{SD}$ and $\mathcal{MAD}$ is still efficient. Therefore, GMS $\mathcal{MAD}$ is even faster than the highly efficient GMSD. GMS $\mathcal{DD}$ is slightly slower than GMSD; however, its improved performance over GMS $\mathcal{SD}$ is noticeable.

%



\section{Conclusion}
\label{conclusion}
Deviation pooling strategies for full reference image quality assessment were analyzed. The mean absolute deviation (MAD) pooling and the standard deviation (SD) pooling strategies were compared on the basis of their effectivity, robustness and efficiency. The computation of MAD is faster than SD, and this may be of high interest for designing more efficient indices. While none of them fully outperformed the others, MAD pooling shows a clear advantage of robustness over SD pooling. Furthermore, for some of the image quality assessment models, a combination of these two pooling strategies results in better performing indices. Considering the experimental results, we highly recommend the use of MAD pooling for different image quality assessment purposes.

{\small
\bibliographystyle{ieee}
\bibliography{egbib}
}

\end{document}